\documentclass{emulateapj}
\usepackage{amssymb}

\begin{document}

\emph{The Astrophysical Journal Letters, 832, L20 (2016)}

\title{MHD-kinetic transition in imbalanced Alfv\'{e}nic turbulence}
\author{Yuriy Voitenko and Johan De Keyser}

\shorttitle{MHD-kinetic turbulence transition}
\shortauthors{Voitenko \& De Keyser}

\begin{abstract}
Alfv\'{e}nic turbulence in space is usually imbalanced: amplitudes of waves
propagating parallel and anti-parallel to the mean magnetic field $\mathbf{B}%
_{0}$ are unequal. It is commonly accepted that the turbulence is driven by
(counter-) collisions between these counter-propagating wave fractions.
Contrary to this, we found a new ion-scale dynamical range of the turbulence
established by (co-) collisions among waves co-propagating in the same
direction along $\mathbf{B}_{0}$. Co-collisions become stronger than
counter-collisions and produce steep non-universal spectra above certain
wavenumber dependent on the imbalance. Spectral indexes of the strong
turbulence vary around $\gtrsim -3$, such that steeper spectra follow larger
imbalances. Intermittency steepens the $-3$ spectra further, up to $-3.7$.
Our theoretical predictions are compatible with steep variable spectra
observed in the solar wind at ion kinetic scales, but further verification
is needed by correlating observed spectra with measured imbalances.
\end{abstract}



\affil{Solar-Terrestrial Centre of Excellence, BIRA-IASB, \\
Ringlaan-3-Avenue Circulaire, B-1180 Brussels, Belgium}

%%%%%%%%%%%%%%%%%%%%%%%%%%%%%%%%%%%%%%%%%%%%%%%%%%%%%%%%%%%%%%%%%%%%%%%%%%%%%%%%%%%%%%%%%%%%%%%%%%%%%%%%%%%%%%%%%%%%%%%%%%%%%%%%%%%%%%%%%%%%%%%%%%%%%%%%%%%%%%%%%%%%%%%%%%%%%%%%%%%%%%%%%%%%%%%%%%%%%%%%%%%%%%%%%%%%%%%%%%%%%%%%%%%%%%%%%%%%%%%%%%%%%%%%%%%%

\section{Introduction}

Theory of strong Alfv\'{e}nic turbulence (Goldreich \& Sridhar 1995)
predicts that the turbulence cascades anisotropically, mainly toward large
perpendicular wavenumbers $k_{\perp }$ (small perpendicular scales $\lambda
_{\perp }=2\pi /k_{\perp }$) across the mean magnetic field, $\mathbf{k}%
_{\perp }\perp \mathbf{B}_{0}$. This prediction has been supported by
observations of the solar-wind turbulence (MacBride \& Smith 2008, and
references therein). With growing $k_{\perp }$ the turbulent fluctuations
become highly anisotropic, $k_{\perp }\gg k_{z}$ ($\mathbf{z}\parallel
\mathbf{B}_{0}$), and their perpendicular scales approach the ion gyroradius
$\rho _{i}$. In this wavenumber range MHD Alfv\'{e}n waves (AWs) transform
into kinetic Alfv\'{e}n waves (KAWs) (Hasegawa \& Chen 1976).

Nonlinear KAW interactions differ significantly from nonlinear AW
interactions and produce different turbulent spectra (Voitenko 1998a,b,
Schekochihin et al. 2009, Voitenko \& De Keyser 2011, Boldyrev \& Perez
2012, and references therein). Consequently, at certain sufficiently large
wavenumber $k_{\bot }=k_{\bot \ast }$ the ion-scale spectral break should
occur where the MHD AW turbulence transforms into the KAW turbulence. Recent
observations of the solar wind turbulence at ion kinetic scales support this
scenario (He et al. 2012, Podesta 2013, Bruno et al. 2014, Roberts et al.
2015). KAW turbulence linked to MHD sources can develop also in solar (Zhao
et al. 2013), terrestrial (Moya et al. 2015, Stawarz et al. 2015), and
Jovian (von Papen et al. 2014) magnetospheres.

Theories of Alfv\'{e}nic turbulence are relatively well developed in
asymptotic MHD ($k_{\perp }^{2}\rho _{i}^{2}\ll 1$) and kinetic ($k_{\perp
}^{2}\rho _{i}^{2}\gg 1$) ranges with perpendicular wavenumber spectra $\sim
k_{\perp }^{-5/3}$ (or $\sim k_{\perp }^{-3/2}$) and $\sim k_{\perp }^{-7/3}$%
, respectively (Goldreich \& Sridhar 1995, Gogoberidze 2007, Schekochihin et
al. 2009, and references therein). The reference cross-field scale
separating MHD and kinetic ranges is the ion gyroradius, $1/k_{\perp \ast
}\sim \rho _{i}$, because finite-$k_{\perp }\rho _{i}$ effects distinguish
KAWs from AWs. However, there is debate on the nature of ion-scale spectral
break $k_{\bot \ast }$ and steep spectra at $k_{\bot }>k_{\bot \ast }$. Ion
gyroradius $\rho _{i}$, ion inertial length $\delta _{i}$, plasma $\beta $,
turbulence amplitude $B/B_{0}$, turbulence anisotropy $k_{\bot \ast
}/k_{z\ast }$, and several their combinations have been suggested as
relevant parameters fixing $k_{\bot \ast }$ (Markovskii et al. 2008, Chen et
al. 2014, Boldyrev et al. 2015).

Solar-wind turbulence is imbalanced - amplitudes of waves propagating from
the Sun $B_{k\left( +\right) }$ are usually larger than amplitudes of
sunward waves $B_{k\left( -\right) }$, Which can affect spectral transport
(e.g. Beresniak \& Lazarian 2008, Gogoberidze \& Voitenko 2016, Yang et al.
2016, and references therein). A common theoretical assumption is that
collisions between these counter-propagating Alfv\'{e}n wave fractions
generate turbulence (Howes \& Nielson 2013, and references therein). In this
Letter, we show that collisions among co-propagating waves (co-collisions
thereafter) at $k_{\perp \ast }<$ $k_{\perp }<$ $1/\rho _{i}$ are stronger
than counter-collisions and establish a new dynamical range of the
turbulence. We shall refer to this as the weakly dispersive range (WDR).
Similarly, we refer to the range $k_{\perp }\rho _{i}>1$,\ where the kinetic
modifications are strong, as the strongly dispersive range (SDR).

\section{Model and basic relations}

Nonlinear dynamic equation for Alfv\'{e}n wave amplitudes, including both
counter- and co-collisions of waves, has been derived by Voitenko (1998a).
Here we construct a semi-phenomenological model for the strong imbalanced
Alfv\'{e}nic turbulence from MHD to kinetic scales using the following
approximation for the nonlinear interaction rate:
\begin{equation}
\gamma _{k\pm }^{\mathrm{NL}}=\frac{2+s}{4\pi }k_{\perp }V_{A}\Delta _{k,s}%
\frac{B_{k\left( \pm s\right) }}{B_{0}},  \label{nir}
\end{equation}%
where the wave velocity mismatch $\delta V_{ks}/V_{A}\equiv $ $\Delta
_{k,s}= $ $\sqrt{1+\left( k_{\perp }\rho _{T}\right) ^{2}}-s$ and magnetic
amplitude $B_{k\left( \pm s\right) }=B_{k\pm }$ for co-collisions ($s=1$)
and $B_{k\left( \pm s\right) }=B_{k\mp }$ for counter-collisions ($s=-1$). (%
\ref{nir}) is obtained from equation (6.3) by Voitenko (1998a) assuming
local interactions and separating dominant (+) and sub-dominant (-) waves
propagating in opposite directions along $\mathbf{B}_{0}\parallel \mathbf{z}$%
. Other definitions are: $\rho _{T}^{2}\simeq $ $\left( 3/4+T_{ez}/T_{i\perp
}\right) \rho _{i}^{2}$ at $k_{\perp }\rho _{i}<1$ and $\rho _{T}^{2}\simeq $
$\left( 1+T_{ez}/T_{i\perp }\right) \rho _{i}^{2}$ at $k_{\perp }\rho _{i}>1$%
, $T_{ez}$ - parallel electron temperature, $T_{i\perp }$- perpendicular ion
temperature, $\rho _{i}=V_{Ti}/\Omega _{i}$ - ion gyroradius, $\Omega _{i}$
- ion gyrofrequency, $V_{Ti}=\sqrt{T_{i\perp }/m_{i}}$ - ion thermal
velocity, $V_{A}=B_{0}/\sqrt{4\pi nm_{i}}$ - Alfv\'{e}n velocity.

A simple phenomenological interpretation of (\ref{nir}) can be given in
terms of colliding waves $1$\ and $2$. The straining rate experienced by
wave $1$\ in the magnetic shear of wave $2$, is proportional not only to the
shear\ $\lambda _{\perp }^{-1}\left( B_{k2}/B_{0}\right) \sim $\ $\left(
2\pi \right) ^{-1}k_{\perp }\left( B_{k2}/B_{0}\right) $, but also to the
relative velocity $V_{\mathrm{ph}1}-sV_{\mathrm{ph}2}$\ defining how fast
the wave $1$\ moves across the shear. Product of these two factors,
accounting for locality $k_{\perp 1}\sim $\ $k_{\perp 2}\sim $\ $k_{\perp }$%
\ and KAW's dispersion $V_{\mathrm{ph}}=$ $V_{A}\sqrt{1+\left( k_{\perp
}\rho _{T}\right) ^{2}}$, gives (\ref{nir}) within a factor of order one.
The key element of (\ref{nir}) that distinguishes co- and counter-collisions
is $\Delta _{k,s}$. Co-collisions ($s=1$) exist only for finite $k_{\perp
}\rho _{T}\neq 0$ making $\Delta _{k,s}\neq 0$ and allowing co-propagating
waves to move with respect to each other undergoing mutual straining.
Counter-collisions ($s=-1$) operate throughout, $\Delta _{k,s}\geq 2$ for
all $k_{\perp }\rho _{T}\geq 0$, as the counter-propagating waves pass
through each other even if they are non-dispersive.

At $k_{\perp }\rho _{T}<1$ (\ref{nir}) reduces to
\begin{equation}
\gamma _{k\pm }^{\mathrm{NL}\left( \uparrow \uparrow \right) }=\frac{1}{2\pi
}\left( k_{\perp }\rho _{T}\right) ^{2}k_{\perp }V_{A}\frac{B_{k\pm }}{B_{0}}%
,  \label{imbal-tail}
\end{equation}%
for co-collisions (superscript $\uparrow \uparrow $), and
\begin{equation}
\gamma _{k\pm }^{\mathrm{NL}\left( \uparrow \downarrow \right) }=\frac{1}{%
2\pi }k_{\perp }V_{A}\frac{B_{k\mp }}{B_{0}}.  \label{imbal-head}
\end{equation}%
for counter-collisions (superscript $\uparrow \downarrow $).

In SDR, $k_{\perp }\rho _{T}>1$, (\ref{nir}) gives
\[
\gamma _{k+}^{\mathrm{NL}\left( \uparrow \downarrow \right) }\approx \frac{1%
}{3}\gamma _{k+}^{\mathrm{NL}\left( \uparrow \uparrow \right) }\approx \frac{%
1}{2\pi }\left( k_{\perp }\rho _{T}\right) k_{\perp }V_{A}\frac{B_{k-}}{B_{0}%
},
\]%
i.e. co-collisions and counter-collisions produce the same scalings.

\section{MHD-kinetic transition and spectra}

In the asymptotic $k_{\perp }\rho _{T}\rightarrow 0$ MHD limit $\gamma
_{k\pm }^{\mathrm{NL}\left( \uparrow \uparrow \right) }\rightarrow 0$ and
the turbulence is driven by counter-collisions, in compliance with Goldreich
\& Sridhar (1995) and many others. The cascade rate in the dominant
component $\gamma _{k+}^{\mathrm{TC}\left( \uparrow \downarrow \right) }\sim
$ $\left( B_{k-}/B_{k+}\right) ^{\mu }\gamma _{k+}^{\mathrm{NL}\left(
\uparrow \downarrow \right) }$, where $\mu =$ $0$, $1/2$, and $1$\ in the
models by Lithwick et al. (2007), Beresniak \& Lazarian (2008), and Chandran
(2008), respectively. For all $\mu $,\ the co-collision rate (\ref%
{imbal-tail}) increases with $k_{\perp }$\ faster than $\gamma _{k+}^{%
\mathrm{TC}\left( \uparrow \downarrow \right) }$\ and the transition occurs
at
\begin{equation}
k_{\perp \ast }\rho _{T}\simeq \left( \frac{B_{k\ast \left( -\right) }}{%
B_{k\ast \left( +\right) }}\right) ^{1/2+\mu }.  \label{trans}
\end{equation}%
Above this wavenumber, the cascade is controlled by kinetic-type
co-collisions.

The turbulence imbalance shifts $k_{\perp \ast }$\ well below $1/\rho _{i}$\
opening window for a new dynamical range WDR $k_{\perp \ast }<$\ $k_{\perp }<
$\ $1/\rho _{i}$. In what follows we consider the most unfavorable case $\mu
=0$\ (Lithwick et al. 2007) with largest $k_{\perp \ast }$\ and narrowest
WDR.

\subsection{Scaling relations}

In the strong turbulence, energy fluxes $\epsilon _{\pm }=$ $\left( \gamma
_{k\pm }^{\mathrm{NL}\left( \uparrow \downarrow \right) }+\gamma _{k\pm }^{%
\mathrm{NL}\left( \uparrow \uparrow \right) }\right) B_{k\pm }^{2}/\left(
4\pi \right) $ can be presented as
\begin{equation}
\epsilon _{\pm }\approx \frac{B_{0}^{2}}{4\pi }\frac{k_{\perp }V_{A}}{4\pi }%
q_{k}\left( \frac{B_{k\mp }}{B_{k\pm }}+p_{k}\right) \left( \frac{B_{k\pm }}{%
B_{0}}\right) ^{3},  \label{e+-}
\end{equation}%
where $q_{k}=\Delta _{k,-1}$ and $p_{k}=3\Delta _{k,1}/\Delta _{k,-1}$ are
regular functions growing with $k_{\perp }$. Using (\ref{e+-}) we express
the fluxes ratio as
\begin{equation}
\frac{\epsilon _{-}}{\epsilon _{+}}=\frac{\left( 1+p_{k}\frac{B_{k-}}{B_{k+}}%
\right) }{\left( \frac{B_{k-}}{B_{k+}}+p_{k}\right) }\left( \frac{B_{k-}}{%
B_{k+}}\right) ^{2}.  \label{e/e}
\end{equation}%
Real solution of this third-order equation for $B_{k-}/B_{k+}$\ is
straightforward but too cumbersome to show explicitly. Denoting it by $b_{k}$%
, we find from (\ref{e+-}) the amplitude scaling $B_{k+}\sim \left[ k_{\perp
}q_{k}\left( b_{k}+p_{k}\right) \right] ^{-3}$\ and spectrum
\begin{equation}
P_{k+}\equiv k_{\perp }^{-1}B_{k+}^{2}\sim k_{\perp }^{-1}\left[ k_{\perp
}q_{k}\left( b_{k}+p_{k}\right) \right] ^{-6}.  \label{spektr}
\end{equation}

At $k_{\perp }\rho _{T}<1$, to the leading order, $p_{k}\approx 0.75\left(
k_{\perp }\rho _{T}\right) ^{2}$, $q_{k}\approx 2$, and the amplitude ratio
\begin{equation}
\frac{B_{k-}}{B_{k+}}\equiv b_{k}\approx \frac{1}{2}\left( \frac{\epsilon
_{-}}{\epsilon _{+}}+\sqrt{\left( \frac{\epsilon _{-}}{\epsilon _{+}}%
+4\left( k_{\perp }\rho _{T}\right) ^{2}\right) \frac{\epsilon _{-}}{%
\epsilon _{+}}}\right) .  \label{b/b1}
\end{equation}%
Depending on $4\left( k_{\perp }\rho _{T}\right) ^{2}\gtrless \epsilon
_{-}/\epsilon _{+}$, the former "MHD" range $k_{\perp }\rho _{T}<1$ splits
into asymptotic MHD range controlled by counter-collisions, and kinetic WDR
controlled by co-collisions.

\textbf{In the asymptotic MHD range} $k_{\perp }\rho _{T}<0.5\sqrt{\epsilon
_{-}/\epsilon _{+}}$ the amplitude ratio (\ref{b/b1}) is $%
B_{k-}/B_{k+}\approx $ $\epsilon _{-}/\epsilon _{+}$ and (\ref{e+-}) gives
the amplitude scaling $B_{k\pm }\sim $ $k_{\perp }^{-1/3}$ and power
spectrum $P_{k\pm }=$ $k_{\perp }^{-1}B_{k\pm }^{2}\sim $ $k_{\perp }^{-5/3}$%
. These scalings reproduce those reported previously.

\textbf{In WDR} $0.5\sqrt{\epsilon _{-}/\epsilon _{+}}<$ $k_{\perp }\rho
_{T}<$ $1$, controlled by co-collisions, the amplitude ratio is $k_{\perp }$%
-dependent:
\begin{equation}
b_{k}\approx \sqrt{\frac{\epsilon _{-}}{\epsilon _{+}}}\left( k_{\perp }\rho
_{T}\right) .  \label{e/e2}
\end{equation}%
Then $B_{k+}\sim $ $k_{\perp }^{-1}$ and spectrum
\begin{equation}
P_{k+}=k_{\perp }^{-1}B_{k+}^{2}\sim k_{\perp }^{-3}.  \label{P2}
\end{equation}%
Subdominant amplitudes $B_{k-}\sim $ $\mathrm{const}$ and $P_{k-}\sim
k_{\perp }^{-1}$.

\textbf{In SDR} $k_{\perp }\rho _{T}>1$ we have $p_{k}\approx 3$, $%
q_{k}\approx \left( k_{\perp }\rho _{T}\right) ^{2}$, then $B_{k\pm }\sim
k_{\perp }^{-2/3}$ and $P_{k\pm }\sim k_{\perp }^{-7/3}$ in both components.

Evolution of (+) waves in WDR disconnects from (-) waves. As the linear
decorrelation rate is $\gamma _{k}^{\mathrm{L}\left( \uparrow \uparrow
\right) }\sim $ $\omega _{k}^{\mathrm{dis}\left( \uparrow \uparrow \right)
}\approx $ $0.5k_{z}V_{A}\left( k_{\perp }\rho _{T}\right) ^{2}$ (dispersive
part of frequency), $k_{z+}\approx \mathrm{const}$ follows from the critical
balance condition $\gamma _{k}^{\mathrm{L}\left( \uparrow \uparrow \right)
}\sim \gamma _{k}^{\mathrm{NL}\left( \uparrow \uparrow \right) }$. Evolution
of parallel scales is thus suppressed in WDR.

\subsection{Non-universal spectra}

If WDR is narrow (it is one order or less in the solar wind), the asymptotic
spectrum $k_{\perp }^{-3}$ can hardly set up. Instead, variable spectra with
indexes approaching $-3$ are expected in WDR. This behavior is observed in
Fig. 1 where the spectra (\ref{spektr}) are plotted without using asymptotic
limits. The spectral indexes in WDR vary $\gtrsim -3$, such that steeper
spectra follow larger imbalances $\epsilon _{+}/\epsilon _{-}$.

\begin{figure}[th]
\figurenum{1} \plotone{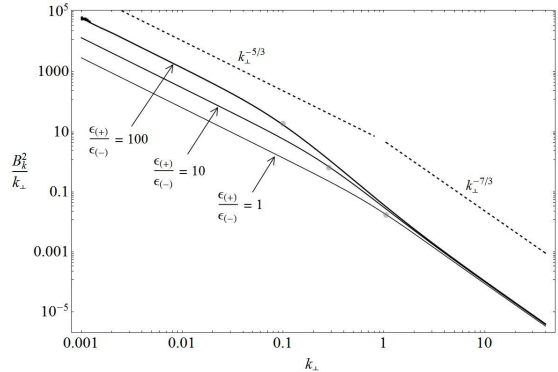}
\caption{Spectra of the dominant (+) component of the strong Alfvenic
turbulence (\protect\ref{spektr}) for different imbalance ratios $\protect%
\epsilon _{\left( +\right) }/\protect\epsilon _{\left( -\right) }$. Grey
dots show breaks $k_{\perp \ast }$ calculated from (\protect\ref{trans}).
Perpendicular wavenumber $k_{\perp }$ is normalized by $\protect\rho _{T}$,
spectral powers are normalized to the same level in SDR kinetic limit.
Asymptotic MHD and SDR spectra -5/3 and -7/3 are shown for reference. }
\label{fig1}
\end{figure}
Spectrum of $-$ waves in WDR is much shallower, $k_{\perp }^{-1}$, which
leads to the convergence of $+$ and $-$ spectra. This effect is seen in Fig.
2.

\begin{figure}[th]
\figurenum{2} \plotone{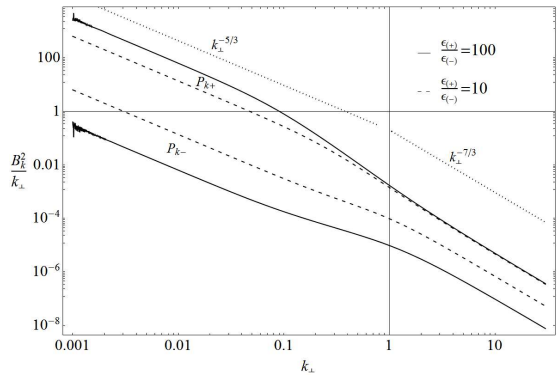}
\caption{Spectra of the dominant (+) and sub-dominant (-) components of the
strong turbulence for two imbalance ratios, $\protect\epsilon _{\left(
+\right) }/\protect\epsilon _{\left( -\right) }$ = 10 (dashed curves) and
100 (solid curves). The spectra converge stronger for larger imbalance.
Other notations as in Fig. 1.}
\label{fig2}
\end{figure}

The imbalance of magnetic amplitudes is shown in Fig. 3. The amplitude ratio
decreases from $B_{k+}/B_{k-}=$ $\epsilon _{+}/\epsilon _{-}$ in the
asymptotic MHD range to $B_{k+}/B_{k-}=$ $\sqrt{\epsilon _{+}/\epsilon _{-}}$
in the asymptotic kinetic range $k_{\perp }\gg $ $\rho _{i}^{-1}$. The
actual drop of the amplitude ratio is larger than the factor $\sqrt{\epsilon
_{+}/\epsilon _{-}}$ because co-collisions are already partially operating
before $k_{\perp \ast }$ and after $\rho _{i}^{-1}$. Say, if the original
imbalance in the asymptotic MHD range is 30, then in the asymptotic kinetic
range above $k_{\perp \ast }$ it drops to about 4, as is seen in Fig. 3
(upper curve).

\begin{figure}[th]
\figurenum{3} \plotone{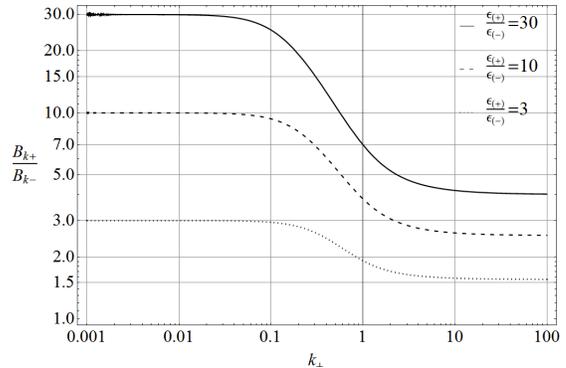}
\caption{Ratio of (+)/(-) magnetic amplitudes for three imbalance ratios, $%
\protect\epsilon _{\left( +\right) }/\protect\epsilon _{\left( -\right) }$ =
30, 10, and 3 from top to bottom. The amplitude ratio decreases
significantly in WDR. }
\label{fig3}
\end{figure}

WDR spectra are affected by intermittency (Boldyrev \& Perez 2012, Zhao et
al. 2016). The modified spectra can be presented as $\tilde{P}_{k}\sim $ $%
k_{\perp }^{-\alpha /3}P_{k}$, where $\alpha =1$ for sheet-like and $\alpha
=2$ for tube-like fluctuations (Zhao et al. 2016). The WDR spectrum $%
P_{k}\sim k_{\perp }^{-3}$ thus steepens to $\tilde{P}_{k}\sim $ $k_{\perp
}^{-3.7}$, close to the steepest spectra reported by Leamon et al. (1999),
Smith et al. (2006), and Sahraoui et al. (2010). Spectra $k_{\perp }^{-4}$
and even steeper can be formed in the weakly turbulent regime (Voitenko
1998b, Galtier \& Meyrand 2015).

\section{Discussion}

Some recent observations are compatible with WDR triggered by imbalance.
Bruno et al. (2014) and Bruno \& Telloni (2015) have reveiled that the
ion-scale spectra are systematically steeper in the faster solar winds and
suggested it may be caused by Alfv\'{e}nicity, i.e. imbalance. Our theory
supports this suggestion and explains why steeper spectra follow larger
imbalances. Chen et al. (2014) have found that at low $\beta $ the spectral
breaks shift to scales larger than $\rho _{i}$\ and associated them with $%
\delta _{i}$, which is hard to explain. We suggest that observed break
scales can be related not to $\delta _{i}$, but to $2\pi /k_{\perp \ast }$\
defined by (\ref{trans}). The required imbalances $B_{k+}/B_{k-}=2.1$\ and $%
3.1$\ in the models $\mu =1$\ and $1/2$\ are realistic; $B_{k+}/B_{k-}=9.8$\
in the model $\mu =0$\ is less realistic. Observed spectral trends (see Fig.
2(a) by Chen et al. (2014) and Fig. 1(b) by Wicks et al. 2011) are the same
as in our Figs. 1 and 2, and agree with other WDR properties. Markovskii et
al. (2007) argued that the break wavenumber decreases with increasing
amplitude at break, which may be caused by the turbulence imbalance (it is
usually larger at larger turbulence level). These observations are
compatible with our theoretical predictions, but they did not measure
imbalances to correlate with spectra. We are not aware of such observations
so far.

Kinetic damping at ion scales has been widely discussed as a possible
barrier for turbulent cascades (see e.g. Leamon et al. 1999, Voitenko \&
Goossens 2004, Wu \& Yang 2007, Podesta et al. 2010, Maneva et al. 2015,
Nariyuki et al. 2014, Cranmer 2014, Passot \& Sulem 2015). Nevertheless,
nearly universal power-law turbulent spectra $k_{\perp }^{-2.8\pm 0.3}$ are
observed in SDR up to electron gyroscales (Alexandrova et al. 2013, and
references therein). Slight deviations from the theoretical spectra ($%
k_{\perp }^{-7/3}$ in strong and $k_{\perp }^{-2.5}$ in weak turbulence) can
be attributed to intermittency (Boldyrev et al. 2012) and damping (Passot \&
Sulem 2015). Actually, Zhao et al. (2016) argued that the damping modifies
the spectral index by $0.1$ only. This suggests that the damping is not so
strong as thought before. In particular, quasi-linear diffusion reduces
velocity-space gradients and wave damping (Voitenko \& Goossens 2006,
Pierrard \& Voitenko 2013), which is supported by observations (He et al.
(2015). We thus focused on nonlinear dynamics ignoring linear damping $%
\gamma _{k}^{\mathrm{L}}$.

Non-universal spectra $k_{\perp }^{-2}$ to $k_{\perp }^{-4}$ observed at $%
k_{\perp }\lesssim $ $1/\rho _{i}$ (Leamon et al. 1999, Smith et al. 2006,
Sahraoui et al. 2010) are much steeper than the theoretical spectrum $%
k_{\perp }^{-5/3}$ formed by counter-collisions. Such strong steepening can
hardly be caused by intermittency or damping without significant nonlinear
modifications. Our theoretical results uphold the dominant role of nonlinear
interactions at ion kinetic scales, where they are strengthened by
co-collisions.

\section{Summary}

We studied the MHD-kinetic transition in strong imbalanced Alfv\'{e}nic
turbulence and found a new dynamical range of the turbulence (WDR) at ion
scales. Its main properties are:

1. The MHD-kinetic transition and spectral break in the imbalanced
turbulence occur at $k_{\perp \ast }$ (\ref{trans}), above which the
turbulence is controlled by kinetic-type co-collisions. For existing models
of the imbalanced MHD turbulence ($\mu =0;1/2;1$) the break $k_{\perp \ast }$
falls well below $1/\rho _{i}$ and a new dynamical range WDR arises at $%
k_{\perp \ast }<$ $k_{\perp }<$ $1/\rho _{i}$.

2. Turbulent cascade is accelerated in WDR and produce steep non-universal
spectra. The spectral index vary from $\gtrsim -2$ to $\lesssim -4$ such
that steeper spectra follow larger imbalances, stronger intermittency, or
weak turbulence.

3. Magnetic amplitude ratio $B_{k\left( +\right) }/B_{k\left( -\right) }$ is
not scale-invariant in WDR decreasing from $\epsilon _{+}/\epsilon _{-}$ to $%
\sqrt{\epsilon _{+}/\epsilon _{-}}$. Similarly, dominant and sub-dominant
spectra converge in WDR.

4. Evolution of the parallel wavenumber and frequency slows down in WDR, and
wavenumber anisotropy grows faster.

Models with $\mu >0$\ (Beresnyak \& Lazarian 2008, Chandran 2008) reproduce
the same spectra as in Figs. 1-2 with significantly lower imbalances $%
B_{k\left( +\right) }/B_{k\left( -\right) }$\ than those required in the $%
\mu =0$ model\ (Lithwick et al. 2007). WDR spectra are steeper than in
nearby MHD and SDR ranges, which results in a double-kink spectral pattern.
This and other properties of WDR are compatible with observations of the
solar-wind turbulence at ion kinetic scales. Applicability of our theory to
solar-wind turbulence needs further verifications by correlating observed
spectra with measured imbalances.

\begin{acknowledgements}
This research was supported by the Belgian Science Policy Office (through
Prodex/Cluster PEA 90316 and IAP Programme project P7/08 CHARM).
%% Results of this paper were presented at the THESOW Conference
%% (20-26 September 2015, Cargese, France) and 2nd THOR Workshop
%% (27-29 September 2016, Barcelona, Spain).
\end{acknowledgements}

%% Literature citations
%% command                        & example result
%% \citet{jones90}|               & Jones et al.\ (1990)
%% \citep{jones90}|               & (Jones et al., 1990)
%% \citep{jones90,jones93}|       & (Jones et al., 1990, 1993)
%% \citep[p.~32]{jones90}|        & (Jones et al., 1990, p.~32)
%% \citep[e.g.,][]{jones90}|      & (e.g., Jones et al., 1990)
%% \citep[e.g.,][p.~32]{jones90}| & (e.g., Jones et al., 1990, p.~32)
%% \citeauthor{jones90}|          & Jones et al.
%% \citeyear{jones90}|            & 1990

\end{document}